\documentclass[aps,twocolumn,pra,superscriptaddress,showpacs,tightenlines]{revtex4-1}
\usepackage{amsmath}
\usepackage{graphicx}
\usepackage{epsfig}
\usepackage{amsfonts}
\usepackage{color}

\begin{document}

\newcommand{\nn}{\nonumber} 
\newcommand{\ms}[1]{\mbox{\scriptsize #1}}
\newcommand{\msi}[1]{\mbox{\scriptsize\textit{#1}}}
\newcommand{\dg}{^\dagger}
\newcommand{\smallfrac}[2]{\mbox{$\frac{#1}{#2}$}}
\newcommand{\ket}[1]{| {#1} \ra}
\newcommand{\bra}[1]{\la {#1} |}
\newcommand{\pfpx}[2]{\frac{\partial #1}{\partial #2}}
\newcommand{\dfdx}[2]{\frac{d #1}{d #2}}
\newcommand{\half}{\smallfrac{1}{2}}
\newcommand{\s}{{\mathcal S}}

\title{All-mechanical quantum noise cancellation for accelerometry: \\ broadband with momentum measurements, narrow band without}

\author{Kurt Jacobs}
\affiliation{U.S. Army Research Laboratory, Computational and Information Sciences Directorate, Adelphi, Maryland 20783, USA}
\affiliation{Department of Physics, University of Massachusetts at Boston, Boston, MA 02125, USA}
\affiliation{Hearne Institute for Theoretical Physics, Louisiana State University, Baton Rouge, LA 70803, USA} 
\author{Nikolas Tezak}
\affiliation{Edward L. Ginzton Laboratory, Stanford University, Stanford, California 94305, USA}
\author{Hideo Mabuchi}
\affiliation{Edward L. Ginzton Laboratory, Stanford University, Stanford, California 94305, USA}
\author{Radhakrishnan Balu}
\affiliation{U.S. Army Research Laboratory, Computational and Information Sciences Directorate, Adelphi, Maryland 20783, USA}

\begin{abstract} 
We show that the ability to make direct measurements of momentum, in addition to the usual direct measurements of position, allows a simple configuration of two identical mechanical oscillators to be used for broadband back-action-free force metrology.  This would eliminate the need for an optical reference oscillator in the scheme of Tsang and Caves [Phys.\ Rev.\ Lett.\ 105, 123601 (2010)], along with its associated disadvantages. We also show that if one is restricted to position measurements alone then two copies of the same two-oscillator configuration can be used for narrow-band back-action-free force metrology. 
\end{abstract} 

\maketitle

\section{Introduction}

Time-varying quantities, known as signals or wave-forms, are detected by monitoring a dynamical system that is affected by the signal. Examples are magnetometry~\cite{Geremia03, Wasilewski10, Vuletic12}, gravity-wave detection~\cite{Caves80, Braginsky80, LIGO13, Miller15}, and accelerometry~\cite{Guzman14}. Because the process involves a measurement on an evolving system, quantum back-action noise is an important limiting factor~\cite{Braginsky80, Caves81}. Nevertheless it is possible, somewhat remarkably, to cancel this intrinsic noise source. Beginning with Unruh in 1982 there have been a number of methods proposed, and some demonstrated, for canceling quantum noise over a range of frequencies~\cite{Unruh82, Bondurant84, Bondurant86, Kimble01, Julsgaard01, Briant03, Caniard07, Tsang10, Wasilewski10, Tsang12c, Wimmer14}. 

Here we consider the most recently proposed scheme for back-action noise cancellation, that by Tsang and Caves~\cite{Tsang10, Tsang12c}. Their scheme not only allows cancellation of quantum noise over all frequencies (broadband), but is conceptually simple and elegant. It requires, in addition to the oscillator that detects the signal (the ``probe'' oscillator), a second ``auxiliary'' oscillator that has identical properties except that it does not feel the signal and its frequency is the negative of that of the probe. The scheme works by having the measuring device coherently measure the sum of the positions of the two oscillators (in the case of an optical oscillator the ``position'' is one of the quadratures). To create an oscillator with an effectively negative frequency Tsang and Caves (TC) used an optical oscillator with a frequency very much higher than that of the mechanical probe, and modulated the measurement of the optical oscillator so that, from the point of view of the measurement, it appeared to have a negative frequency. This modulation is achieved by modulating the strength of the interaction between the optical oscillator and the measuring device, and is a fairly standard ``frequency conversion'' technique that is straightforward to implement with optics. The modulation only produces the desired effect, however, if the frequency of the auxiliary, $\omega_{\ms{a}}$, is much larger than that of the probe, $\nu$, because it generates additional terms in the output spectrum that appear at $\pm 2\omega_{\ms{a}}$. These interfere with the measurement process if they overlap with the spectrum of the probe oscillator. 

While elegant the scheme of Tsang and Caves has two drawbacks. The first is that, in order to measure the sum of the position of the probe and a quadrature of the optical oscillator a parametric amplifier must be used to engineer the required interaction with the optical oscillator. The second is that perfect noise cancellation is only achieved when the damping rates of the probe and auxiliary are the same, something that is not easy to achieve with an optical auxiliary. 

The drawbacks of TC noise cancellation would be eliminated if it were possible to use an auxiliary that felt the same force as the probe and had the same frequency. In that case a mechanical oscillator could be used as the auxiliary, which would eliminate the need for a parametric amplifier and make it much easier to achieve identical damping rates. Here we consider how this could be done. 

We begin by showing that if one has the ability to directly measure any quadrature of the mechanical motion, then a small modification of the TC scheme allows one to use an auxiliary that has the same frequency and feels the same signal as the probe. This scheme is conceptually simple and makes it clear that one can perform perfect frequency conversion using a linear interaction so long as one can measure any quadrature (any linear observable). By ``perfect'' frequency conversion we mean an interaction with an oscillator such that the frequency of the oscillator appears to be different from its true frequency, and without the introduction of any spurious sidebands at multiples of the latter.  

Measuring an arbitrary quadrature of a mechanical oscillator may not be easy, however. An arbitrary mechanical quadrature is an arbitrary superposition of the position and momentum. While measuring position is straightforward, as far as we are aware there are only two methods that have to-date been proposed for making direct measurements of momentum, and neither have been demonstrated in the quantum regime. The first of these methods, by Greywall~\textit{et al.},  is to use the magnetic field generated by electrons, since this field is proportional to the electrons' velocity~\cite{Greywall94}. To achieve broadband noise-cancellation our method requires that we measure a time-varying superposition of the momentum and position of one oscillator along with the position of second. An issue with the scheme of Greywall~\textit{et al.} is that electrical current involves dissipation, and because of this it is not obvious how to extend the scheme to measure a coherent superposition of momentum and position. The second scheme by Benetov and Blencowe is more recent, and involves an effective interaction with the momentum of an oscillator that can be obtained using a quantum point-contact~\cite{Benetov12}. While this scheme indicates that it may well be possible to make measurements of a superposition of one quadrature of one oscillator and another quadrature of another, we do not attempt to develop any specific schemes to do so here. Nevertheless, the fact that such measurements could be used for all-mechanical broadband back-action-free accelerometry may provide a motivation for doing so. 

Since techniques for the direct measurement of momentum are not yet well-developed, we also examine whether noise cancellation can be achieved with two identical mechanical oscillators and with position measurement alone. We show that this is possible if one uses two pairs of oscillators rather than a single pair, although because the frequency conversion is imperfect the scheme is restricted to narrow-band detection. By ``narrow-band detection'' we mean that the bandwidth of the force detection must be significantly less than the frequency of the mechanical oscillators. 

This paper is laid out as follows. In Section~\ref{mom} we describe the Tsang-Caves scheme for back-action noise cancellation, and show that it can be performed with two identical mechanical oscillators if one has the ability to measure a time-dependent arbitrary quadrature of each oscillator. In Section~\ref{possec} we show how modulating a position measurement can be used to obtain a similar result, but does not allow broad-band detection. In Section~\ref{imp} we describe a simple opto-mechanical (or electro-mechanical) method for implementing the schemes presented in Section~\ref{possec}. In this section we also discuss the physical parameters required to beat the standard quantum limit. We finish with a brief conclusion in Section~\ref{conc}. 

\section{Broadband noise cancellation with momentum measurement}
\label{mom}

We must first describe how TC noise-cancellation works. To do this all we need are the Heisenberg equations of motion that describe a Harmonic oscillator whose position is continuously measured. These equations are 
\begin{eqnarray}
 \dot{x} & = & \hspace{2.8mm} \Omega p , \\ 
 \dot{p} & = & -\Omega x + \sqrt{8k} \xi(t),  
\end{eqnarray} 
in which $x$ and $p$ are dimensionless versions of the position and momentum of the oscillator, $\Omega$ is the angular frequency of the oscillator, and $\xi(t)$ is a quantum white noise source with correlation function $\langle \xi(t)\xi(t')\rangle = \delta(t'-t)$~\cite{Jacobs14}. The dimensionless position and momentum are given respectively by $x = b + b^\dagger$ and $p = -i(b-b^\dagger)$, where $b$ is the annihilation operator for the oscillator. The parameter $k$ is determined by the rate at which the measurement extracts information about the position. It is related to the noise on the stream of measurement results, $r(t)$, by $r(t) = \langle x(t) \rangle + z(t)/\sqrt{8k}$ where $z(t)$ is white noise with correlation functions $\langle z(t)z(t')\rangle = \delta(t'-t)$ and $\langle z(t)\xi(t')\rangle = 0$. 

To realize TC noise cancellation, we consider having two such oscillators with respective positions $x_1$ and $x_2$, where the second has frequency equal to $-\Omega$. If we measure $X_+ = x_1 + x_2$, then because we are measuring a coherent sum of the two positions, the two oscillators are driven by identical back-action noise. If we drive the positive oscillator with a force $f_1(t)$ then the equations of motion for the two oscillators are 
\begin{align}
  \dot{x}_1 & =  \hspace{2.8mm}  \Omega  p_1, \;\;\; \;  \dot{p}_1  = - \Omega  x_1 + \sqrt{8k} \xi (t)  + f_1(t), \label{m1} \\ 
  \dot{x}_2 & =  -\Omega  p_2,  \;\;\; \;  \dot{p}_2 = \hspace{3mm} \Omega  x_2 + \sqrt{8k} \xi (t) .   \label{m2} 
\end{align} 
If we consider $P_- = p_1 - p_2$ then from the above equations we can obtain immediately $\dot{X}_+ = \Omega P_-$ and $\dot{P}_- = - \Omega X_+ + f_1(t)$. Because $P_-$ is the difference between $p_1$ and $p_2$, and because the back-action noise is common to both, this noise cancels for the motion of $P_-$. Thus $X_+$ and $P_-$ constitute an oscillator that experiences no back-action noise, while still feeling the force $f_1(t)$. This is TC quantum noise cancellation. If we drive the negative oscillator with a force $f_2(t)$, then the noise-free oscillator feels the difference of the forces,  $f_1(t) - f_2(t)$. 

We now note that the variables $X_- = x_1 - x_2$ and $P_+ = p_1 + p_2$ also form an oscillator, in that when there are no driving forces or measurements we have $\dot{X}_- = \Omega P_+$ and $\dot{P}_+ = -\Omega  \tilde{X}_-$. What is more, if we measure $X_- = x_1 - x_2$ then $p_2$ experiences the \textit{negative} of the back-action force that drives $p_1$. Because $p_1$ and $p_2$ experience opposite back-action noise, for the variable $P_+ = p_1 + p_2$ this noise cancels so that 
\begin{align}
 \dot{X}_- &  =  \hspace{2.8mm} \Omega P_+ ,  \\ 
  \dot{P}_+ &  =  -  \Omega  X_-   + f_1(t) + f_2(t). 
\end{align} 
where $f_1(t)$ is the force driving the probe and $f_2(t)$ is that driving the auxiliary. That is, a measurement of $X_- = x_1 - x_2$ realizes a back-action free measurement of the sum of the forces on the two oscillators. Since we can measure the sum of two forces in this way, if both oscillators experience the same force $f = f_1 = f_2$ we could use this method to measure this force if we had such a positive/negative pair. We could do this using two identical oscillators if we could measure them in such a way that the frequency of one of them was effectively negated. 


To create an oscillator with an effectively negative frequency, first consider making a measurement of the position $x$ of an oscillator with frequency $\nu$. If we look at the dynamics in the Heisenberg picture then the measured observable is 
\begin{align}
   x(t) = x_0 \cos(\nu t) + p_0 \sin (\nu t) = a_0 e^{-i\nu t} + a_0^\dagger e^{i\nu t} , 
\end{align} 
where $x_0 = a_0 + a_0^\dagger$ and $p_0 = -i(a_0 - a_0^\dagger)$ are constant operators. We can cancel this time-dependence completely by instead measuring an observable that is rotating in the ``opposite direction'' at the same frequency. That is, if we measure 
\begin{align}
   y_{-\nu}(t) \equiv x \cos(\nu t) - p \sin (\nu t) = a e^{i\nu t} + a^\dagger e^{-i\nu t} , 
\end{align} 
then the result is a measurement of the time-independent quantity $x_0$. Taking this further, if we measure an observable rotating at $-\nu \pm \Omega$, then an oscillator with frequency $\nu$ appears to have frequency $\pm\Omega$. That is, the negative oscillator in Eq.(\ref{m2}) is equivalent to a measurement of 
\begin{align}
   \tilde{x}(t) = y_{-\nu-\Omega}(t) = a e^{i(\nu+\Omega)t} + a^\dagger e^{-i(\nu+\Omega) t}  
\end{align} 
on an oscillator with frequency $\nu$. To obtain an oscillator with frequency $-\nu$ from one with frequency $\nu$ we just have to measure $y_{-2\nu}(t)$. Note that to do this we must be able to make a measurement of a superposition of $x$ and $p$. 

The above measurement procedure performs perfect frequency conversion. But we are not quite done yet, because we need to examine how the force driving the oscillator appears in the measurement signal. In short, since the force is driving a real oscillator with frequency $\nu$, and the measurement sees an oscillator with frequency $-\nu$, the force must appear transformed because it would produce a different signal if it were really driving a negative frequency oscillator. We can determine this transformation easily by deriving the equation of motion for $y_{-2\nu}(t)$ from the equations of motion for $x$ and $p$. By differentiating $y \equiv y_{-2\nu}(t)$ we find that 
\begin{eqnarray}
 \dot{y} & = &  - \nu p_y - \sin(2\nu t) f(t),  \label{effneg1} \\ 
 \dot{p_y} & = & \hspace{2.8mm} \nu y + \cos(2\nu t) f(t) ,   \label{effneg2}
\end{eqnarray} 
in which 
\begin{eqnarray}
 p_y & \equiv & x \sin(2\nu t) + p \cos (2\nu t) . 
\end{eqnarray} 
The effective negative oscillator is thus driven by a transformed version of the force. We need to see how the force appears in the output signal so that we can process this signal appropriately to back-out $f$. To do this it is simplest to work in frequency space, and this requires that we add damping to our oscillators so that they have a well-defined steady-state. 

When the damping rate of an oscillator is small compared to its frequency then the usual linear frictional damping force, namely $\dot{p} = -\gamma p$ for some damping rate $\gamma$, transforms approximately into damping that is symmetric in $x$ and $p$, and as a result can be modeled by a simple Markovian master equation~\cite{Jacobs14}. The equations of motion for a positive oscillator under symmetric damping are 
\begin{align}
   \dot{x} & = -\frac{\gamma}{2} x + \nu p+ \sqrt{\gamma}v_{\ms{p}} (t) , \\
   \dot{p} & = -\frac{\gamma}{2} p - \nu x + \sqrt{\gamma}v_{\ms{x}} (t), 
\end{align} 
where $v_x(t)$ and $v_p(t)$ are white noise sources with the correlation functions 
\begin{align}
   \langle v_x(t) v_x(t')\rangle & = \langle v_p(t) v_p(t')\rangle = (2 n_T +1)\delta(t-t') \\ 
   \langle v_x(t) v_p(t')\rangle & = 0.  
\end{align} 
The parameter $n_T$ is the average number of thermal phonons in the oscillator at the ambient temperature $T$~\footnote{The average number of phonons at temperature T is $n_T = [\exp(\hbar\nu/kT)-1]^{-1}$, where $\nu$ is the oscillator frequency and $k$ is Boltzmann's constant~\cite{Jacobs14}.}. 

By converting the above equations of motion to frequency space we can determine how the Fourier transform of the force appears in the Fourier transform of the position. Adding driving terms to the equations of motion to give 
\begin{align}
   \dot{x} & = -\frac{\gamma}{2} x + \nu p + \sqrt{\gamma}v_{\ms{p}} (t) + s_{\ms{x}}(t), \\
   \dot{p} & = -\frac{\gamma}{2} p - \nu x + \sqrt{\gamma}v_{\ms{x}} (t) + s_{\ms{p}}(t), 
\end{align} 
we find that these terms appear in the position as 
\begin{align}
    x_f(\omega) & = \frac{ \nu S_{\ms{p}}(\omega) + (\gamma/2 - i\omega)S_{\ms{x}}(\omega)}{ G(\omega)}  , 
\end{align} 
where $x_f(\omega)$ is the total contribution of the driving terms to the Fourier transform of the oscillator position, $S_{\ms{x}}(\omega)$ and $S_{\ms{p}}(\omega)$ are the Fourier transforms of the driving terms, and we have defined 
\begin{align}
   G(\omega) \equiv \left( \gamma/2- i\omega \right)^2 + \nu^2  . 
\end{align} 

Applying the above result to the oscillator given by Eqs.(\ref{effneg1}) and (\ref{effneg2}) we find that the force appears in the position $y$ as 
\begin{align} 
    y_f(\omega) & = \frac{F(\omega + \nu)}{A(\omega - \nu)} - \frac{F(\omega - \nu)}{A(\omega+\nu)},  \label{ysig}
\end{align} 
where $A(s) \equiv s + i\gamma/2$ and we have used the fact that $G(\omega) = A(\omega + \nu) A(\omega - \nu).$ 

To obtain noise cancellation we need to measure a sum of $y$ and the coordinate $x$ for a positive oscillator with frequency $\nu$. Since we do not need to perform any frequency conversion for this measurement, the force appears in the signal of the measurement of $x$ as 
\begin{align} 
   x_f(\omega) & = \frac{\nu F(\omega)}{ G(\omega)} . \label{xsig} 
\end{align} 
Adding the signals in Eqs.(\ref{ysig}) and (\ref{xsig}) together to give $z_f = x_f + y_f$ the combined output signal is 
\begin{align} 
    z_f(\omega) & = -  \frac{F(\omega - \nu)}{A(\omega + \nu)} + \frac{\nu F(\omega)}{G(\omega)} + \frac{F(\omega + \nu)}{A(\omega-\nu)} . 
\end{align} 
The measured signal therefore contains components of the force spectrum at three different frequencies. There are various ways that we can extract the force from this signal. We first note that if we substitute $\omega_n = (n+1)\nu + \omega$ into $z_f$ we obtain  
\begin{align}
   F_{n}   & = - a_n z_f(\omega_n)  + \frac{a_n}{b_n} \nu F_{n+1} - \frac{a_n}{c_n} F_{n+2}, 
\end{align} 
where $a_n$, $b_n$, and $c_n$ are functions of $\omega_n$ and we have defined $F_n = F(n\nu + \omega)$. The above expression is a recursion relation for $F(\omega)$. Since the transfer function of the oscillators has a width of $\gamma$ and scales as $1/\nu$ for $\nu \gg \gamma$, the contribution of a term containing $F_{n}$ decreases approximately as $1/n$.  We can therefore use the above recursion relation and truncate it at a value of $n$ that gives the desired accuracy. 

A much neater expression for the force can be obtained by using a second pair of oscillators identical to the first pair except that instead of measuring $y(t)$ for the second oscillator we measure the observable that lags (or leads) $y(t)$ by $\pi/2$. That is, we measure 
\begin{align} 
    y'(t) = x \sin(\nu t) + p \cos (\nu t) , 
\end{align} 
which gives the output signal 
\begin{align} 
    z'_f(\omega) & = i  \frac{F(\omega - \nu)}{A(\omega + \nu)} + \frac{\nu F(\omega)}{G(\omega)} + i \frac{F(\omega + \nu)}{A(\omega-\nu)} . 
\end{align} 
Adding together the signals $z_f$ and $z'_f$ we have  
\begin{align} 
  \frac{F(\omega)}{G(\omega)}  & =  \frac{z_f(\omega) - i  z'_f(\omega)}{\nu(1-i)} - 2 \frac{ F(\omega + \nu)}{(1-i) \nu A(\omega-\nu)} , 
\end{align} 
which provides the simple recursion relation 
\begin{equation}
   F_n = \alpha_n - \beta_n F_{n+1} 
\end{equation}
with $F_n = F(\omega_n) = F(\omega + n\nu)$ and 
\begin{align} 
        \alpha_n & =  G(\omega_n) \frac{z_f(\omega_n) - i  z'_f(\omega_n)}{(1-i)\nu} \\
        \beta_n & = \left( \frac{2}{1-i} \right) \frac{A(\omega_{n+1})}{\nu} . 
\end{align} 
The resulting exact expression for the force is 
\begin{align} 
      F(\omega) = \sum_{n=0}^\infty (-1)^n \alpha_n \left(  \prod_{k=0}^{n-1} \beta_k \right) . 
\end{align} 
Each subsequent term in the sum makes a smaller contribution to $F(\omega)$, so the sum can be truncated when sufficient accuracy is achieved. 

%
%

%

\section{Narrowband noise cancellation with position measurement} 
\label{possec}

We now examine how close we can get to the above noise cancellation scheme if we are limited to position measurements alone. This means that the interaction with each of the mechanical oscillators is restricted to the form $H_{\ms{int}} = \hbar \lambda B x$ where $x$ is the mechanical position, $B$ is a quadrature of the probe oscillator, and $\lambda$ is a rate constant. Given this interaction it is well-known that we can still perform an approximate frequency conversion by modulating the interaction rate $\lambda$ (see, for example~\cite{Tian09, Liao14, Jacobs14}). Let us modulate the coupling as 
\begin{equation}
   \lambda(t) = \lambda_0 \left\{ \cos([\omega_a - \nu + \Omega]t) + \cos([\omega_a + \nu - \Omega]t) \right\} , 
\end{equation}
where $\omega_a$ is the frequency of the probe oscillator, $\nu$ is the that of the mechanical oscillator, and $\Omega$ is the frequency that wish the mechanical oscillator to appear to have. We will assume that the probe oscillator is optical and thus has a much greater frequency that the mechanics. If we denote the annihilation operator for the probe by $b$ so that $A = b + b^\dagger$ and that of the mechanical oscillator by $a$, and move into the interaction picture with respect to both oscillators, the Hamiltonian becomes 
\begin{eqnarray}
   H_{\ms{int}}^{\ms{I}} & =  & \hbar \lambda_0 A [ x \cos(\Omega t) + p \sin(\Omega t) ]  \nonumber \\ 
                                    & + & \hbar \lambda_0 A [ x \cos([\Omega-2\nu] t) + p \sin([\Omega-2\nu] t) ] , 
\end{eqnarray}
with some additional terms that oscillate at the much higher frequency $\omega_a$. In the first interaction term the mechanical oscillator appears to be rotating at frequency $\Omega$ instead of frequency $\nu$, and if this were the only interaction term we would have perfect frequency conversion. In the second interaction term the mechanical oscillator appears to be oscillating at $2\nu-\Omega$. This second term will in general interfere with our noise cancellation scheme. It produces a signal in the output of the probe that is shifted from that of the first by $2(\nu - \Omega)$, and will also generate back-action noise driving the mechanics that is shifted up by this  frequency. 

%
%
%
%
%
%
%
%

The additional fictitious oscillator at $2(\nu - \Omega)$ ceases to be a problem if its spectrum is well-separated from the spectrum of the fictitious oscillator with frequency $\Omega$. This will be the case if the bandwidth of the mechanics, which is essentially the mechanical damping rate, as well as the frequency $\Omega$ are both much less than $2\nu$. Thus the modulation scheme for performing approximate frequency conversion will allow us to construct a back-action cancellation scheme if the bandwidth we wish to measure is small compared to the mechanical frequency $\nu$. 

In the cancellation scheme above we started with two mechanical oscillators with frequency $\nu$ and used a measurement of a quadrature rotating at $-2\nu$ to obtain a oscillator with effective frequency $-\nu$. Now using our modulated position measurement the frequency of the effective oscillator that we produce, $\Omega$, must be considerably smaller than $\nu$. We therefore start with two oscillators at frequency $\nu$ and use the modulation technique to give one of them an effective frequency of $\Omega$ and the other an effective frequency of $-\Omega$. This gives us the required positive/negative pair.  

Because the bandwidth $\gamma$ passed by the mechanical oscillators is much smaller than its central frequency $\nu$ the positive and negative parts of the spectrum of the oscillator (the spectrum of the motion of the oscillator's position coordinate) are well-separated and can be treated separately when considering transformations on this spectrum. Recall that the negative frequency part of the Fourier transform, $F(\omega)$, of a real signal is the complex conjugate of the positive part: $F(-\omega) = F^*(\omega)$. The fact that the positive and negative parts are initially well-separated provides some simplification in the analysis; a measurement that shifts the frequency of an oscillator from $\nu$ down to $\Omega$ shifts the positive part of the Fourier transform of its motion down from $\nu$ to $\Omega$ and the negative frequency part up from $-\nu$ to $-\Omega$. We will denote the positive part of a Fourier transform $F(\omega)$ by adding the superscript ``pos'' to give $F^{\ms{pos}}(\omega)$. 

We now use the modulation method above to measure two mechanical oscillators with frequency $\nu$ so that they appear to have, respectively, the frequencies $\pm\Omega$. The effective equations of motion for these oscillators, minus the damping and quantum back-action, are then
\begin{eqnarray}
 \dot{y_{\pm}} & = &  \pm\Omega \, p_{\pm} - \sin(\nu\mp\Omega t) f(t),  \label{effosc1} \\ 
 \dot{p_{\pm}} & = &  \mp\Omega \, y_{\pm} + \cos(\nu\mp\Omega t) f(t) ,   \label{effosc2}
\end{eqnarray} 
with $p_\pm \equiv x \sin(\pm\Omega  t) + p \cos (\pm\Omega  t)$. To obtain quantum noise cancellation we measure the observable 
\begin{align} 
   z = y_{+} + y_{-} , 
\end{align} 
and the force appears in the positive frequency part of this signal as
\begin{widetext}
\begin{eqnarray} 
     z_f^{\ms{pos}}(\omega) & = & \frac{\gamma/2 - i [\omega-\Omega]}{2 G(\omega)}  \Bigl[ F^{\ms{pos}}(\omega + [\nu - \Omega]) + F^{\ms{neg}}(\omega - [\nu + \Omega]) + F^{\ms{pos}}(\omega + [\nu + \Omega]) + F^{\ms{neg}}(\omega - [\nu - \Omega]) \Bigr]  ,  \label{eqzfnarr}
\end{eqnarray} 
\end{widetext}
where $G(\omega) = (\gamma/2 - i\omega)^2 + \Omega^2$. The negative part of the spectrum of the force, $F^{\ms{neg}}(\omega)$, appears here for the following reason. The modulation that generates the effective positive oscillator with frequency $+\Omega$ shifts the positive frequency part of the force down from $+\nu$ to $+\Omega$ and the negative part from $-\nu$ to $-\Omega$. However, the modulation that generates the effective negative oscillator moves the negative part sitting at $-\nu$ up to $+\Omega$, and similarly the positive part at $+\nu$ down to $-\Omega$. Thus a contribution from the negative part of the force spectrum appears in the positive part of the output signal, and vice versa. 

\subsubsection{Case I: Bandwidth $\gamma \ll \Omega$}

\begin{figure}[t] 
\includegraphics[width=0.82\hsize]{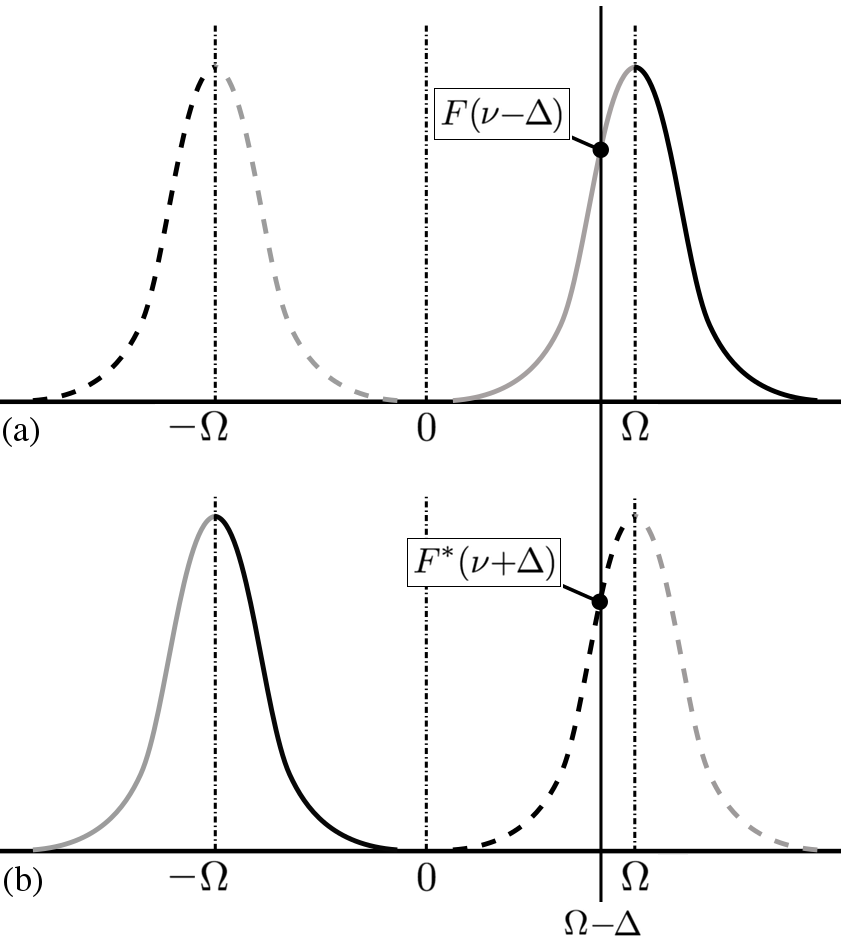}
\caption{The Fourier transforms of the output signals from the effective oscillators are shown in (a) (frequency $\Omega$) and (b) (frequency $-\Omega$). In (a) the ``hump'' on the positive frequency side has been shifted down from $\omega = \nu$ to $\omega=\Omega$, and that on the negative frequency side has been shifted up from $\omega = -\nu$ to $\omega = -\Omega$. In (b) the hump on the left is the same hump as that on the right in (a), and vice versa.  Since each hump is not symmetric, we have drawn the right and left sides of the hump that comes from $\omega = \nu$ with black and grey solid lines, respectively. We have similarly depicted the hump that comes from $-\nu$ with black and grey dashed lines, so that a dashed line of a given color denotes the complex conjugate of the solid line with the same color.} 
\label{fig1} 
\end{figure}

We can now distinguish two cases. If we chose the bandwidth of the oscillators so that it is much smaller than $\Omega$, then the spectrum centered at $-\Omega$ does not overlap with that which is centered at $\Omega$. That is, the spectrum centered at $-\Omega$ does not encroach on the positive part, and so the last two terms in the above expression for $z_f^{\ms{pos}}(\omega)$ drop out. This situation is illustrated in Fig.~\ref{fig1}. In Fig.~\ref{fig1}a we show the spectrum of the signal coming from the positive oscillator and in Fig.~\ref{fig1}b the spectrum coming from the negative oscillator. If we write $\omega = \Omega + \Delta$ then we have 
\begin{eqnarray} 
     \frac{z_f^{\ms{pos}}(\Omega + \Delta)}{B(\Omega + \Delta)} & = & F^{\ms{pos}}(\nu + \Delta) + F^{\ms{neg}}(-\nu + \Delta)   , 
\end{eqnarray} 
where 
\begin{eqnarray}
   B(\omega) = \frac{\gamma/2 - i (\omega - \Omega)}{2 G(\omega)} . 
\end{eqnarray}
Noting that the relationship between the positive and negative parts of the spectrum is $F(-\omega) = F^*(\omega)$ we can also write this as 
\begin{eqnarray} 
      \frac{z_f^{\ms{pos}}(\Omega + \Delta)}{B(\Omega + \Delta)}  & = &  F^{\ms{pos}}(\nu + \Delta) + F^{\ms{pos}*}(\nu - \Delta)  . 
\end{eqnarray} 
To determine the spectrum of the force from the output signal we need only disentangle $F^{\ms{pos}}(\nu + \Delta)$ from $F^{\ms{pos}*}(\nu - \Delta)$. We can do this, as we did in the previous section, by having another pair of oscillators in which we measure quadratures that are out of phase with $y_{\pm}$ by $\pi/2$. The result of changing the phase of the measured observables is that a different linear combination of the positive and negative spectra appear in the output signal, due to the fact that we are changing the phase relationship between the force and the effective oscillators. 

If we denote by $y_{\pm}^{(\pi/2)}$ the quadratures that respectively lag $y_{\pm}$ by $\pi/2$, and we measure the observable 
\begin{align} 
   \tilde{z} = y_{+}^{(\pi/2)} + y_{-}^{(\pi/2)} , 
\end{align} 
then the force appears in the output of this measurement as 
\begin{eqnarray} 
    \frac{\tilde{z}_f^{\ms{pos}}(\Omega + \Delta)}{ i B(\Omega + \Delta)}  & = & F^{\ms{pos}}(\nu + \Delta]) -  F^{\ms{pos}*}(\nu - \Delta)   . 
\end{eqnarray} 
We can therefore obtain the force as 
\begin{eqnarray} 
    F^{\ms{pos}}(\nu + \Delta]) =   \frac{z_f^{\ms{pos}}(\Omega + \Delta) - i \tilde{z}_f^{\ms{pos}}(\Omega + \Delta)}{2 B(\Omega + \Delta)} . 
\end{eqnarray} 

\subsubsection{Case II: Bandwidth $\gamma \sim \Omega$}

\begin{figure}[t] 
\includegraphics[width=0.82\hsize]{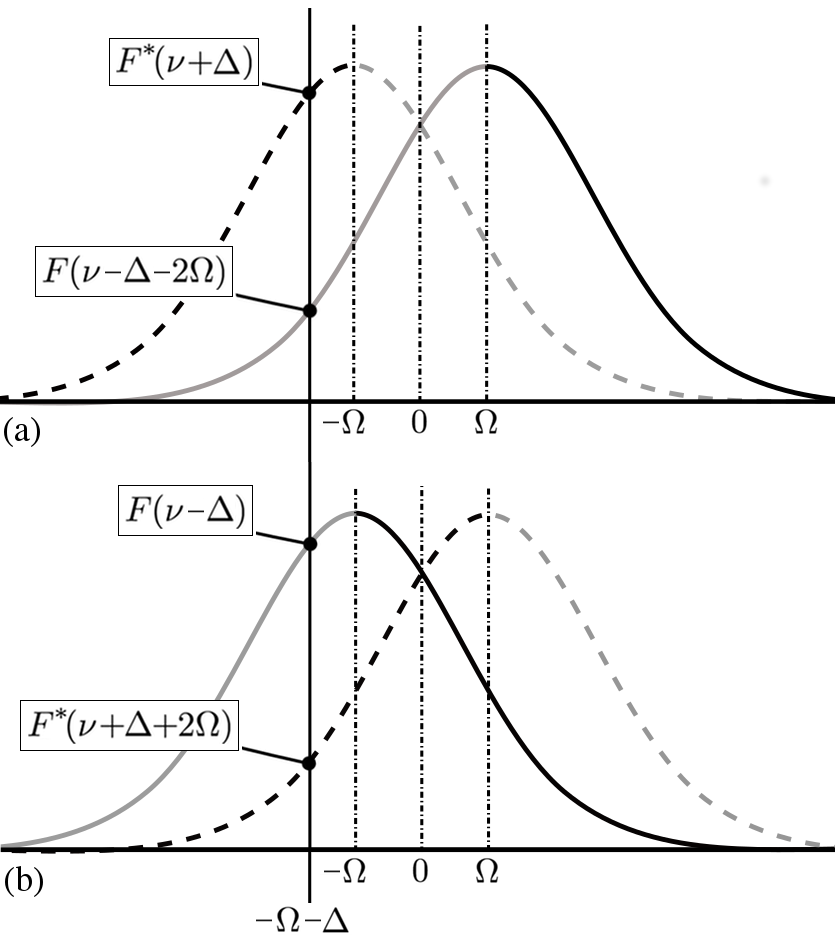}
\caption{The Fourier transforms of the output signals from the effective oscillators are shown in (a) (frequency $\Omega$) and (b) (frequency $-\Omega$) in the situation in which the humps that have been shifted down from $\omega = \nu$ (and up from $\omega = -\nu$) overlap. As in Fig.~\ref{fig1} the dashed lines represent the (reflected) complex conjugates of the solid lines with the same color.} 
\label{fig2}
\end{figure}

Ideally we would like to detect the broadest possible bandwidth and so we now consider the case in which $\gamma$ is not much smaller than $\Omega$. In this case the positive and negative parts of the spectrum overlap in the output signal, as depicted in Fig.~\ref{fig2}. It is now useful to introduce the following compact notation: 
\begin{align} 
    F_{n}^\pm & \equiv F(\nu + n\Omega \pm \Delta) ,  \\ 
   Z_{n}^\pm & \equiv  \frac{ 2 z_f^{\ms{pos}}([n+1]\Omega \pm \Delta)}{ B([n+1]\Omega \pm \Delta)}, \\ 
    \tilde{Z}_{n}^\pm & \equiv  \frac{- 2 i \tilde{z}_f^{\ms{pos}}([n+1]\Omega \pm \Delta)}{B([n+1]\Omega \pm \Delta)} . 
\end{align} 
With these definitions Eq.(\ref{eqzfnarr}) becomes
\begin{eqnarray} 
    Z_{n}^+ & = & \frac{1}{2}\left(  F_{n}^+ + F^{-*}_{-n} + F^+_{(n+2)} + F^{-*}_{-(n+2)}  \right),  \label{eqzfnarr2}
\end{eqnarray} 
and the measurement of the quadrature that lags $Z$ by $\pi/2$ gives the signal 
\begin{eqnarray} 
     \tilde{Z}_{n}^+ & = & \frac{1}{2}\left( F_{n}^+ - F^{-*}_{-n} + F^+_{(n+2)} - F^{-*}_{-(n+2)} \right) , 
\end{eqnarray} 
so that 
\begin{eqnarray} 
   Z_{n}^+  +  \tilde{Z}_{n}^+ & = &  F_{n}^+  + F^+_{(n+2)}  . 
\end{eqnarray} 
Solving this recursion relation gives us   
\begin{align}
    F(\nu + \Delta) =  \sum_{n = 0}^\infty (-1)^n \left( Z_{2n}^+  +  \tilde{Z}_{2n}^+ \right) .  \label{bbd1}
\end{align}
As $n$ increases the values of $Z_{2n}^+$ and $\tilde{Z}_{2n}^+$ decrease as $1/n$ so that the series can be truncated when the desired level of accuracy is reached. Since the transfer function is Lorentzian with width $\gamma$, and the sample points are separated by $2\Omega$, it is the ratio $r \equiv \gamma/\Omega$ that determines the number of values, $N$, that are required in the series. Specifically, if we demand an error $\varepsilon \ll 1$, then we can truncate the sum in Eq.(\ref{bbd1}) at $N \approx r/\varepsilon$. 

\section{Optomechanical implementation for narrow-band detection}
\label{imp}

\begin{figure}[t] 
\includegraphics[width=1\hsize]{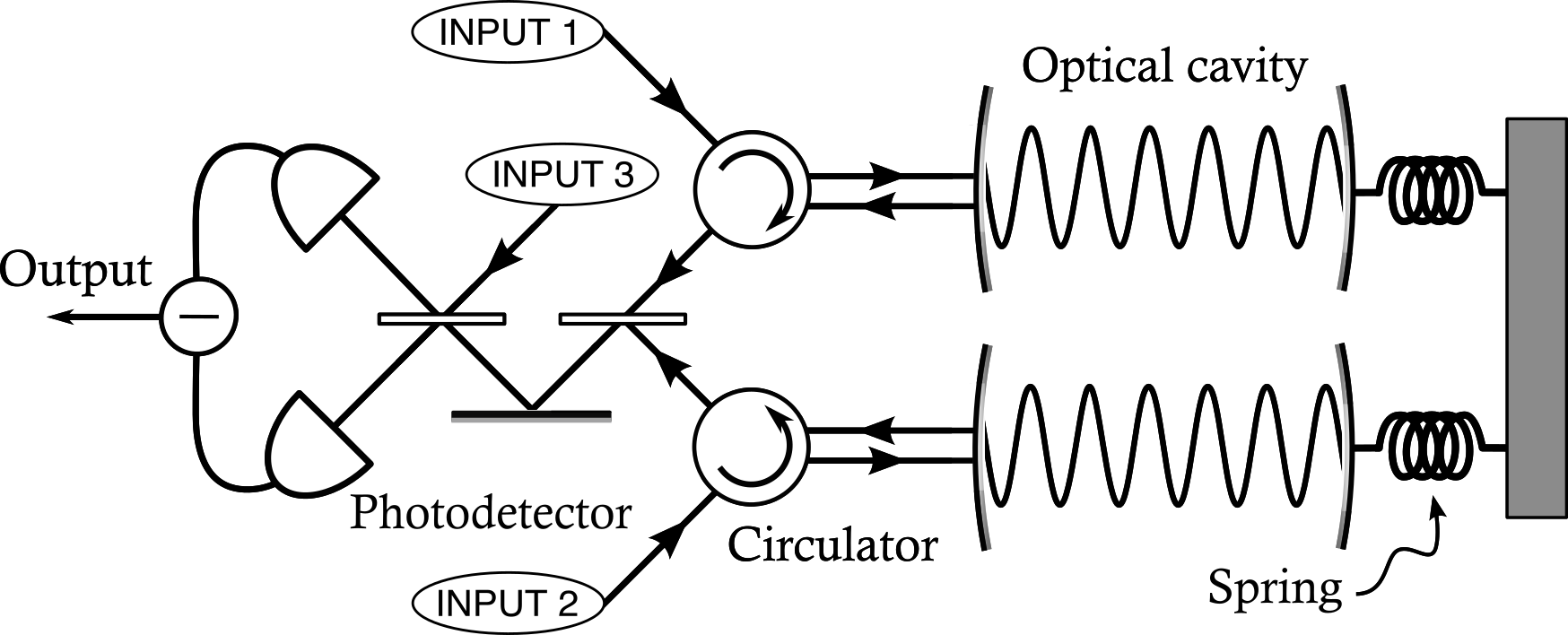}
\caption{A diagram of a physical implementation of our force metrology scheme for a single pair of mechanical oscillators. Laser inputs 1 and 2 drive the optical/superconducting cavities, one being modulated at $\Omega$ and the other at $-\Omega$. Input 3 is the local oscillator for the homodyne detection, which is performed after interfering the cavity outputs on a 50/50 beamsplitter. In the simplest scenario, two of these oscillator-pair configurations are used, where in the second the laser inputs 1 and 2 are shifted by $\pi/2$ from those of the first. Each configuration then provides one of the quadratures of the applied force.} 
\label{fig3}
\end{figure}

In Fig.~\ref{fig3} we depict an implementation of the narrow-band noise-cancellation scheme presented in Section~\ref{possec}. While this implementation is shown using optical cavities, superconducting nano-electromechanics can also be used and is identical from a theoretical point of view~\cite{Jacobs14}. In Fig.~\ref{fig3} each of the two mechanical oscillators is coupled to an optical oscillator in the usual opto-mechanical manner~\cite{Clerk08}: each mechanical oscillator forms one of the end-mirrors of the corresponding optical or superconducting cavity. The cavities have a resonance frequency $\omega_{\ms{c}}$ that is much higher than the mechanical resonance $\nu$. To enable measurement of a mechanical quadrature rotating at $-\nu \pm\Omega$, so as to provide effective oscillators with frequencies $\pm\Omega$, the cavities are driven on resonance, and the driving laser is amplitude modulated at $\nu\pm\Omega$. The term in the Hamiltonian that describes this driving is 
\begin{equation}
    H_{\ms{d}} =  i\hbar \beta \cos([\nu\pm\Omega]t) (a e^{-i\omega_{\ms{c}}t} - a^\dagger e^{i\omega_{\ms{c}}t}) , 
\end{equation}
where $a$ is the mode operator for the cavity mode and $\beta$ is proportional to the amplitude of the laser. This combination of driving frequency and amplitude modulation implements the modulation scheme described in Section~\ref{possec}. 

Once we eliminate the interaction terms oscillating with frequencies greater than or equal to $2\nu$ the Hamiltonian that couples a cavity to one of the mechanical oscillators is 
\begin{eqnarray} 
    H & = &  \hbar \nu b^\dagger b + \hbar g (c + c^\dagger) y_\pm(t) ,  
\end{eqnarray}
where 
\begin{align}
y_\pm(t) = b e^{i (\nu \pm \Omega) t} + b^\dagger e^{-i (\nu \pm \Omega) t}, 
\end{align}
$b$ is the mechanical mode operator, $g = \alpha g_0$ is the effective coupling rate, $|\alpha|^2$ is the steady-state number of photons in the cavity mode, and $g_0$ is the opto-mechanical coupling strength~\cite{Law95, Jacobs14}. The mode operator $c = a - \alpha$ is a shifted version of the cavity mode operator. We have written the Hamiltonian $H$ in the interaction picture with respect to the oscillation of the optical mode at frequency $\omega_{\ms{c}}$. Because we will measure the output of the cavity at the optical frequency it is $H$ that correctly describes the observed dynamics. Because the observables $y_\pm(t)$ is coupled to the cavity amplitude quadrature, $x_a = a + a^\dagger$, a measurement of the phase quadrature reads them out. By shifting the phase of the modulation we can alternatively measure phase-shifted versions of $y_\pm(t)$ as required by the schemes in Section~\ref{possec}. 

To measure a coherent superposition of the quadratures of two mechanical oscillators we interfere the outputs of the two cavities at a beamsplitter before measuring the phase, and this completes the implementation shown in Fig.~\ref{fig3}. This scheme is well-suited to demonstration with current experimental systems~\cite{Schliesser09, Chan11, Palomaki13}. 

\subsubsection*{Beating the standard quantum limit}

We now discuss the physical parameters required to beat the standard quantum limit. To do so one must examine the key noise sources for force measurement. These are i) thermal noise, ii) classical measurement noise also referred to as ``measurement inefficiency''~\cite{Jacobs14}, and iii) the quantum back-action noise that is eliminated by the present scheme. The back-action noise and thermal noise drive the oscillator in the same way as the force signal being measured, so they can be compared very simply. They are both filtered through the transfer function of the mechanical oscillators which, at resonance, means dividing their noise powers by $\gamma^2$. The fundamental minimum measurement noise, which is white noise added to the output signal $\tilde{x}(t)$ has power $S_{\ms{M}}^{\ms{min}}(\omega) = 1/(8k)$, the inverse of the back-action. Any noise above this level is due to classical noise in the detection system, including noise on the driving laser. The measurement efficiency is defined by $\eta = S_{\ms{M}}^{\ms{min}}/S_{\ms{M}}$ where $S_{\ms{M}}$ is the actual measurement noise.

We will present all noise powers in the dimensionless units of $\tilde{p}$, since these are easily converted to units of real force by multiplying by $\hbar\nu m/2$, where $m$ is the mass of the mechanical oscillators. Because we are measuring a combination of variables of two oscillators the contributions from the various noise sources are slightly different than those for a single oscillator.  The total noise on the output signal in the above dimensionless units is 
\begin{equation} 
   S_{\ms{out}} = \frac{1}{8\eta k} + \frac{4(2n_T + 1)}{\gamma} + \frac{4\langle \mbox{Re}[F(\nu)]^2 \rangle}{\gamma^2} + \frac{8k}{\gamma^2} ,  
\end{equation}
where we have used a measurement of the real part of $F(\nu)$ as our example. The first term in $S_{\ms{out}}$ is the measurement noise, the second is thermal noise and the fourth is the back-action noise that would normally appear but which is eliminated by the quantum noise cancellation. The measurement rate $k = 2g/\kappa$ where $\kappa$ is the damping rate of the cavities. This damping rate gives the frequency response of the measurement so we usually want to have $\kappa \gtrsim \nu$. At zero temperature the thermal noise reduces to the zero-point motion of the oscillators. 

If we define the point at which the measurement is limited by back-action noise as the point at which the back-action noise is equal to the thermal noise, then the criterion for demonstrating cancellation is $k \geq \gamma (n_T + 1/2)$. In terms of the cavity photon number $|\alpha|^2$ and opto-mechanical coupling constant $g_0$ this criterion is $\alpha g_0 \geq \gamma \kappa (2n_T + 1)/4$. 

\section{Conclusion}
\label{conc}
 
We have shown that it is possible to realize quantum noise cancellation for force detection using two identical mechanical oscillators. We have shown that if one can directly measure the momentum as well as the position of a mechanical oscillator then two identical oscillators can be used for back-action free force detection across all frequencies, so-called ``broadband'' detection. While we have not presented a method to make such momentum measurements, this result provides a motivation for doing so. 

At least as far as we have been able to determine, it is not possible to achieve broadband detection with two identical oscillators if one only has position measurement at ones disposal. The reason for this is that one cannot then measure an arbitrary quadrature of the mechanics, and as a result an attempt to perform frequency conversion --- required to create an oscillator with an effectively negative frequency --- introduces additional noise in various places in the spectrum. We have shown that it is nevertheless possible to use two identical mechanical oscillators to make back-action free measurements for a narrow band of frequencies about the mechanical frequency. This is possible because modulating a position measurement allows one to realize approximate frequency conversion for a bandwidth that is small compared to the mechanical frequency. 

%


\end{document}